\documentclass[Physsubmission, Phys]{SciPost}
\hypersetup{
    colorlinks,
    linkcolor={red!50!black},
    citecolor={blue!50!black},
    urlcolor={blue!80!black}
}

\usepackage[bitstream-charter]{mathdesign}
\usepackage{slashed}
\usepackage{tikz-feynman}

\usepackage{tikz-feynman}

\DeclareSymbolFont{usualmathcal}{OMS}{cmsy}{m}{n}
\DeclareSymbolFontAlphabet{\mathcal}{usualmathcal}

\def \e  {\mathop{\rm e}\nolimits}

\newcommand \vev [1] {\langle{#1}\rangle}

\newcommand \ket [1] {|{#1}\rangle}
\newcommand \bra [1] {\langle {#1}|}

\begin{document}

\begin{center}{\Large \textbf{The pion-photon transition form factor at two loops in QCD\\
}}\end{center}

% TODO: write the author list here. Use initials + surname format.
% Separate subsequent authors by a comma, omit comma at the end of the list.
% Mark the corresponding author with a superscript *.
\begin{center}
Jing Gao\textsuperscript{1,2,3,4},
Tobias Huber\textsuperscript{2$\star$},
Yao Ji\textsuperscript{5,2} and
Yu-Ming Wang\textsuperscript{1}
\end{center}

\begin{center}
{\bf 1} School of Physics, Nankai University, Weijin Road 94, 300071 Tianjin, China
\\
{\bf 2} Naturwissenschaftlich-Technische Fakult\"at, Universit\"at Siegen, \\ Walter-Flex-Str.~3, 57068 Siegen, Germany
\\
{\bf 3} Institute of High Energy Physics, CAS, P.O. Box 918(4) Beijing 100049, China
\\
{\bf 4} School of Physics, University of Chinese Academy of Sciences, Beijing 100049, China
\\
{\bf 5} Physik Department T31, James-Franck-Stra\ss e 1, Technische Universit\"at M\"unchen, D-85748 Garching, Germany\\
$\star$ \href{mailto:huber@physik.uni-siegen.de}{huber@physik.uni-siegen.de}%\\
%SI-HEP-2021-28,~P3H-21-083,~TUM-HEP-1369/21
\end{center}

\vspace*{-12pt}

\begin{center}
\today
\end{center}

\vspace*{-12pt}

% For convenience during refereeing (optional),
% you can turn on line numbers by uncommenting the next line:
%\linenumbers

\definecolor{palegray}{gray}{0.95}
\begin{center}
\colorbox{palegray}{
  \begin{tabular}{rr}
  \begin{minipage}{0.1\textwidth}
    \includegraphics[width=35mm]{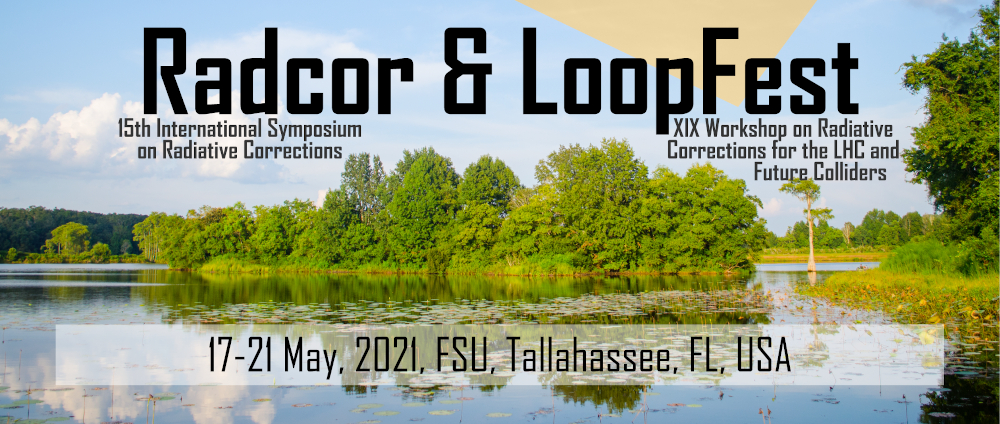}
  \end{minipage}
  &
  \begin{minipage}{0.85\textwidth}
    \begin{center}
    {\it 15th International Symposium on Radiative Corrections: \\Applications of Quantum Field Theory to Phenomenology,}\\
    {\it FSU, Tallahasse, FL, USA, 17-21 May 2021} \\
    \doi{10.21468/SciPostPhysProc.?}\\
    \end{center}
  \end{minipage}
\end{tabular}
}
\end{center}

\vspace*{-24pt}

\section*{Abstract}
{\bf
%The abstract is in boldface and should fit in 8 lines.
We report on the fully analytic calculation of the leading-power contribution
to the photon-pion transition form factor $\gamma \, \gamma^{\ast} \to \pi^0$
at two loops in QCD. The applied techniques are based on hard-collinear factorization,
together with modern multi-loop methods. We focus both, on the technical details
such as the treatment of evanescent operators, and the phenomenological implications.
Our results indicate that the two-loop correction is numerically comparable
to the one-loop effect in the same kinematic domain. We also demonstrate that our
results will play a key role in disentangling various models for the twist-two pion
distribution amplitude thanks to the envisaged precision at Belle II.
}

% TODO: include a table of contents
\vspace{10pt}
\noindent\rule{\textwidth}{1pt}
\tableofcontents\thispagestyle{fancy}
\noindent\rule{\textwidth}{1pt}
\vspace{10pt}

\section{Introduction}
\label{sec:intro}
% TODO: write your article here.
Hard exclusive processes play a prominent role in exploring the strong interaction dynamics of hadronic reactions in the framework of QCD. In this context one of the simplest exclusive matrix elements is the pion-photon transition form factor $F_{\gamma \pi} (Q^2)$ which appears in the process $\gamma \gamma^\ast \to \pi^0$. At large momentum transfer, it serves for testing theoretical predictions based upon perturbative QCD factorization.

\begin{figure*}[t]
\begin{center}
\includegraphics[width=6.5cm]{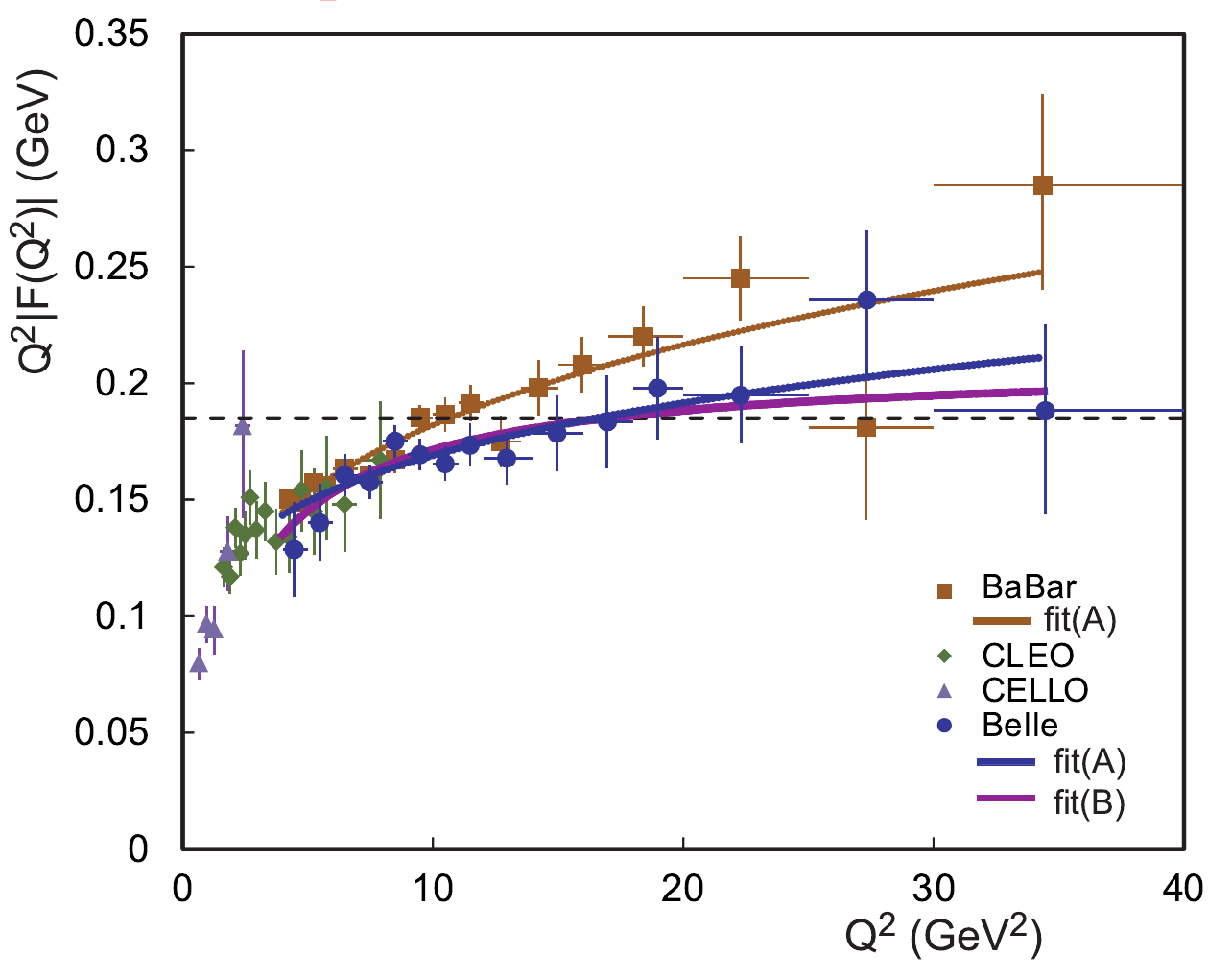}
\end{center}

\vspace*{-10pt}

\caption{The $\gamma \gamma^\ast \to \pi^0$ transition form factor multiplied by
$Q^2$. The dashed line indicates the asymptotic behaviour of the form factor. Taken from~\cite{Uehara:2012ag}. \label{fig:scaling}}
\end{figure*}

Experimentally, the pion-photon transition form factor (TFF) with one on-shell and one off-shell photon can be extracted from measurements of the differential $e^{+} \, e^{-} \to e^{+} \, e^{-} \, \pi^{0}$ cross section~\cite{Gronberg:1997fj,Aubert:2009mc,Uehara:2012ag}. A measurement of BaBar in 2009~\cite{Aubert:2009mc} reported on an unexpected scaling violation of the TFF at large $Q^2$ (see Fig.~\ref{fig:scaling}) which triggered quite some interest in the community. A subsequent analysis at Belle~\cite{Uehara:2012ag} did not find the pronounced increase of the pion TFF in the high-$Q^2$ region, resulting in a moderate overall tension. The large amount of data that will be accumulated at Belle~II will eventually clarify the situation. 

On the theory side the TFF at large momentum transfer is expanded in powers of $\Lambda_{\rm QCD}^2 / Q^2$, and at leading power (LP) it is expressed as a convolution of the perturbatively calculable hard coefficient function (CF) with the twist-two pion light-cone distribution amplitude (LCDA) in accordance with the hard-collinear factorization theorem~\cite{Lepage:1980fj}. While the hard CF has been studied to next-to-leading order (NLO)~\cite{delAguila:1981nk,Braaten:1982yp,Kadantseva:1985kb,Wang:2017ijn} and in the large-$\beta_0$ approximation at next-to-next-to-leading order (NNLO)~\cite{Melic:2001wb}, the full NNLO QCD correction was missing until recently, and we report on its analytic computation in the present article. The pion distribution amplitude, on the other hand, is the non-perturbative object in the factorization formula, which thanks to its universality is of great importance also for other processes such as the semileptonic or nonleptonic decays of $B$-mesons.

The full analytic NNLO QCD prediction of the TFF $F_{\gamma \pi} (Q^2)$ in $\gamma \, \gamma^{\ast} \to \pi^0$ was first put forward in~\cite{Gao:2021iqq}, with a parallel computation appearing in~\cite{Braun:2021grd} and the two results being in full agreement with each other. Our computation takes advantage of the soft-collinear effective theory (SCET) factorization program. To this end, we evaluate an appropriate bare QCD matrix element at ${\cal O}(\alpha_s^2)$ using modern multi-loop techniques, and implement the ultraviolet (UV) renormalization and infrared (IR) subtraction, including the proper treatment of the emerging evanescent operator, together with the subtleties arising from the $\gamma_5$ ambiguity of dimensional regularization. Furthermore, we analyse the numerical impact of the two-loop correction to the TFF, and compare different models for the twist-two pion LCDA to current experimental data. While current data still leaves room for interpretation, confronting the obtained theory predictions with the forthcoming precision of Belle II measurements will allow to distinguish between different LCDA models.

%%%%%%%%%%%%%%%%%%%%%%%%%%%%%%%%%%%%%%%%%%%%%%%%%%%%%%%%%%%%%%%
%%%%%%%%%%%%%%%%%%%%%%%%%%%%%%%%%%%%%%%%%%%%%%%%%%%%%%%%%%%%%%%

%%%%%%%%%%%%%%%%%%%%%%%%%%%%%%%%%%%%%%%%%%%%%%%%%%%%%%%%%%%%%%%%%%%%%%%%%%%%%%%%%%%%%%%%%%%%%%%%%%%%
%%%%%%%%%%%%%%%%%%%%%%%%%%%%%%%%%%%%%%%%%%%%%%%%%%%%%%%%%%%%%%%%%%%%%%%%%%%%%%%%%%%%%%%%%%%%%%%%%%%%

\section{The pion-photon transition form factor}
\label{sec:tff}

We start by setting up the theory framework for establishing the hard-collinear factorization formula
of the transition form factor $F_{\gamma \pi} (Q^2)$, which is defined in terms of the matrix element of the electromagnetic current
between an on-shell photon with momentum $p^{\prime}$ and a pion with momentum $p$,
\begin{align}
\langle \pi(p) | j_{\mu}^{\rm em}  | \gamma (p^{\prime}) \rangle
= g_{\rm em}^2 \epsilon_{\mu \nu \alpha \beta} q^{\alpha}  p^{\beta}  \epsilon^{\nu}(p^{\prime})
F_{\gamma \pi} (Q^2) \, .
\end{align}
Here $q=p-p^{\prime}$ is the transfer momentum and $Q^2=-q^2$. $\epsilon^{\nu}(p^{\prime})$ is the polarization vector of the on-shell photon, and $e_q$ denotes the electric charge of the quark field in units of the positron charge.
For later convenience we also introduce two light-like vectors $n_{\mu}$ and $\bar n_{\mu}$
satisfying $n^2=\bar n^2=0$ and $n \cdot \bar n=2$. They allow for the definition of the perpendicular component of any four-vector via
$a^\mu = (n \cdot a) \, \bar n^{\mu}/2 + (\bar n \cdot a) \, n^{\mu}/2 + a_\perp^\mu$. The kinematics at leading power can then be taken as $p_{\mu} = (\bar n \cdot p) /2 \, n_{\mu}$ and $p^{\prime}_{\mu} = (n \cdot p) /2 \, \bar n_{\mu}$, which entails the scaling $(\bar n \cdot p) \sim (n \cdot p^{\prime}) \sim {\cal O}(\sqrt{Q^2})$.

Applying the hard-collinear factorization theorem results in the following LP contribution
to the $\gamma \, \gamma^{\ast} \to \pi^0$  form factor
\begin{align}
F^{\rm LP}_{\gamma\pi}(Q^2)&=\frac{(e_u^2-e_d^2)f_\pi}{\sqrt{2}\,Q^2}
\int^1_0dx\,T_2(x,Q^2,\mu)\,\phi_\pi(x,\mu) \equiv \, \frac{(e_u^2-e_d^2)f_\pi}{\sqrt{2}\,Q^2}
T_2(x,Q^2,\mu)\otimes \phi_\pi(x,\mu)\, .
\label{eq:factorization}
\end{align}
Here $T_2$ is the hard coefficient function which can be expanded perturbatively
in the form (similarly for any other partonic quantity)
\begin{align}
T_2=\sum_{\ell=0}^\infty  a_s^{\ell} \, T_2^{(\ell)} \,,
\qquad a_s\equiv\frac{\alpha_s}{4\pi}  \,.
\label{eq:loopexpansion}
\end{align}
The non-perturbative object in the factorization formula \eqref{eq:factorization} is the twist-two pion LCDA $\phi_\pi(x,\mu)$, which is defined by the renormalized  matrix element on the light-cone
\begin{align}
\bra{\pi(p)}\,[\bar q(z\bar n)[z\bar n,0]\gamma_\mu\gamma_5 q(0)]_R\,\ket{0}
=-i f_\pi p_\mu\int^1_0dx\,\e^{i x z \bar n \cdot p} \, \phi_\pi(x,\mu)\, .
\label{def:pion-LCDA}
\end{align}
$[z\bar n,0]$ is the short-hand notation for the Wilson line which renders the non-local matrix element gauge invariant.

We evaluate the hard coefficient function by inspecting the following correlation function
\begin{align}
    &\frac{g_{\rm em}^2 e_q^2}{2\,\bar n\cdot p}\,\Pi_{\mu\nu} = i\int \!\! d^4z \e^{-iq\cdot z}
    \times \bra{\bar q(\bar xp) q(x p)}
    {\rm T} \{ j^{\rm em}_\mu(z), \, j^{\rm em}_\nu(0) \}\ket{0} \,,
\end{align}
which can be parameterized by the two form factors for the bilinear quark currents
with the spin structures \cite{Wang:2017ijn}
\begin{align}
 \Gamma_{A}^{\mu\nu} &= \gamma_{\perp}^{\mu} \, \slashed{\bar n} \, \gamma_{\perp}^{\nu} \;, &
 \Gamma_{B}^{\mu\nu} &= \gamma_{\perp}^{\nu} \, \slashed{\bar n} \,  \gamma_{\perp}^{\mu}  \;.
\end{align}
We will devote the next section to the two-loop calculation of the bare matrix element $\Pi_{\mu\nu}^{(2)}$, and subsequently derive the master formula for the hard coefficient function $T_2$.
%
%%%%%%%%%%%%%%%%%%%%%%%%%%%%%%%%%%%%%%%%%%%%%%%%%%%%%%%%%%%%%%%%%%%%%%%%%%%%%%%%%%%%%%%%%%%%%%%%%%%%
%%%%%%%%%%%%%%%%%%%%%%%%%%%%%%%%%%%%%%%%%%%%%%%%%%%%%%%%%%%%%%%%%%%%%%%%%%%%%%%%%%%%%%%%%%%%%%%%%%%%

\section{Two-loop calculation}

The techniques that we apply during the calculation of the bare two-loop amplitude 
have become standard in multi-loop computations. 

%%%%%%%%%%%%%%%%%%%%%%%%%%%%%%%%%%%%%%%%%%%%%%%%%%%%%%%%%%%%%%%
\begin{figure*}[t]
\includegraphics[width=0.21\textwidth]{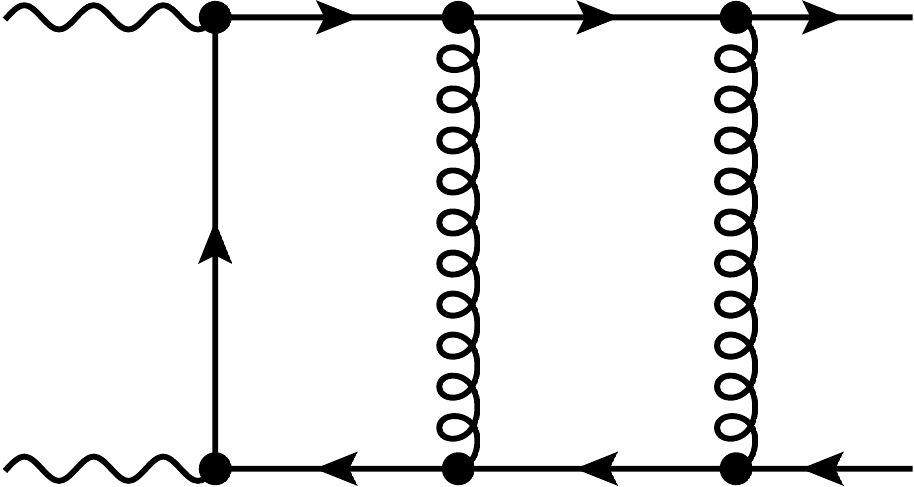}\hspace*{10pt}
\includegraphics[width=0.21\textwidth]{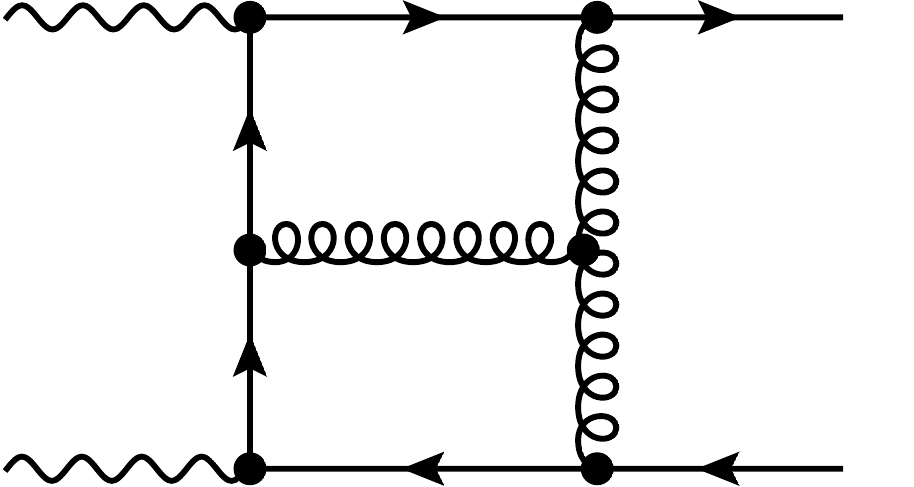}\hspace*{10pt}
\includegraphics[width=0.21\textwidth]{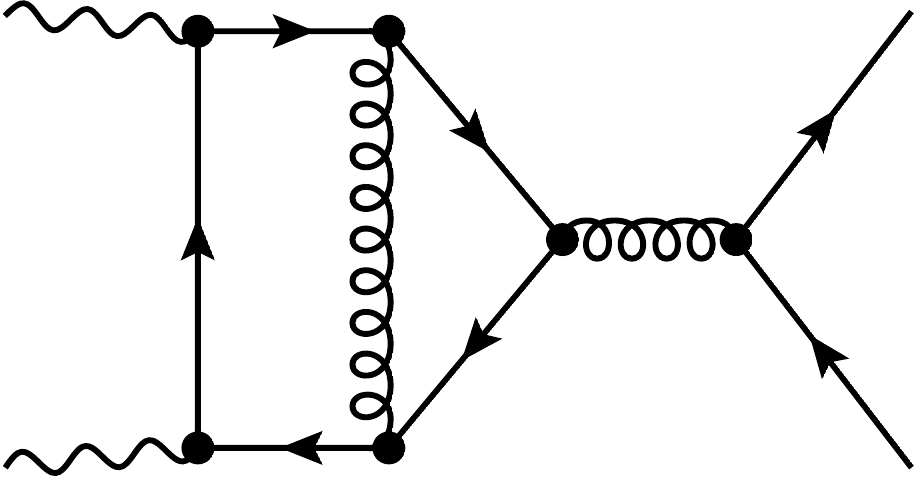}\hspace*{10pt}
\includegraphics[width=0.21\textwidth]{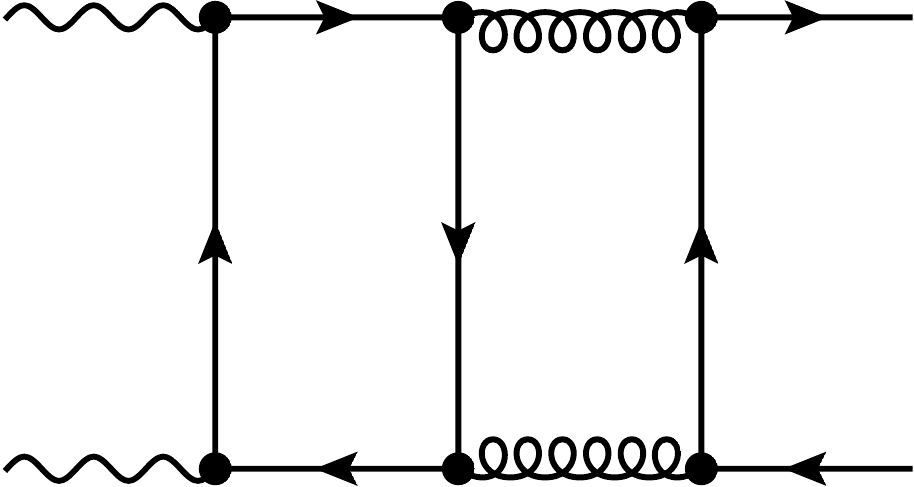}
\caption{Sample Feynman diagrams. }
\label{fig:diagrams}
\end{figure*}
%%%%%%%%%%%%%%%%%%%%%%%%%%%%%%%%%%%%%%%%%%%%%%%%%%%%%%%%%%%%%%%

We generate the Feynman diagrams in two ways,
with {\tt Feynarts}~\cite{Hahn:2000kx}
and by means of an in-house routine. Selected diagrams are shown in Fig.~\ref{fig:diagrams}.
The number of diagrams gets reduced by the fact that certain diagrams have color factor zero
(e.g.\ the third one in Fig.~\ref{fig:diagrams}), vanish due to the Furry theorem and/or
represent a flavor-singlet contribution (e.g.\ the last one in Fig.~\ref{fig:diagrams}).
What remains is a total of $42$ diagrams (plus the ones with the two photons interchanged),
which we compute using dimensional regularization with $D=4-2\epsilon$ to simultaneously regulate UV and IR divergences.

After the Dirac and tensor reduction we are left with two-loop scalar integrals
which we further process with the {\tt Mathematica} version of {\tt FIRE}~\cite{Smirnov:2008iw}, an implementation of
Laporta's algorithm~\cite{Laporta:1996mq,Laporta:2001dd} based on integration-by-parts identities~\cite{Chetyrkin:1981qh,Tkachov:1981wb}.
In addition, we exploit the fact that two of the quark momenta are parallel to each other, $p_1 \equiv x p \propto (1-x) p \equiv \bar x p = p_2$, which yields additional relations between integrals based on momentum conservation, which in turn enable us to arrive at the minimal set of master integrals.
It is also worth mentioning that at this stage, additional Dirac structures besides $\Gamma_{A,B}^{\mu\nu}$
disappear from the sum of all diagrams, making the QED Ward identity and charge symmetry manifest and providing an important check of our calculation.

\begin{figure}[t]
\centering
\scalebox{1.2}{
\scalebox{0.5}{\begin{tabular}{cccc}
\scalebox{1.30}{\feynmandiagram [layered layout, small, horizontal=a to b] {
% Draw the top and bottom lines
i1 [particle=\(p^\prime\)]
--  [fermion] a %[dot]
--  b %[dot]
--  c %[dot]
--  [fermion] f1 [particle=\(p_1\)],
i2 [particle=\(p_2\)]
-- [anti fermion] d %[dot]
--  e %[dot]
--  f %[dot]
-- [anti fermion] f2 [particle=\(q\)],
% Draw the two internal fermion lines
{ [same layer] a -- d },
{ [same layer] b -- e},
{ [same layer] c -- f},
};}&
\scalebox{1.18}{\begin{tikzpicture}
\begin{feynman}
\vertex (a) {$q$};
\vertex[right=1cm of a] (b);
\vertex[right=1cm of b] (ch1);
\vertex[above=0.7cm of ch1] (c1);
\vertex[above=0.5cm of c1] (o1);
\vertex[right=1cm of c1] (c2);
\vertex[right=0.5cm of c2] (o2) {$p_1$};
\vertex[below=0.7cm of ch1] (d1);
\vertex[right=1cm of d1] (d2);
\vertex[right=0.5cm of d2] (o3) {$p^\prime$};
%more auxiliary vertices
\vertex[above=0.05cm of ch1] (p1);
\vertex[below=0.05cm of ch1] (p2);
\vertex[right=0.54cm of p1] (q1);
\vertex[right=0.46cm of p2] (q2);
\diagram* {
{
(a) -- [fermion, small] (b),
(b) -- (c1) -- (c2)  -- [fermion, small] (o2),
(b) -- (d1) -- (d2)  -- [anti fermion, small] (o3),
(c1) -- (d2),(c1) -- [fermion, small, edge label=\(p_2\)] (o1),
(d1) -- (q2), (c2) -- (q1)
}
};
\end{feynman}
\end{tikzpicture}}&
\scalebox{1.18}{\begin{tikzpicture}
\begin{feynman}
\vertex (a) {$q$};
\vertex[right=1cm of a] (b);
\vertex[right=1cm of b] (ch1);
\vertex[above=0.7cm of ch1] (c1);
\vertex[right=1cm of c1] (c2);
\vertex[right=0.5cm of c2] (o2) {$p_1$};
\vertex[below=0.7cm of ch1] (d1);
\vertex[right=1cm of d1] (d2);
\vertex[right=0.5cm of d2] (o3) {$p^\prime-p_2$};
%more auxiliary vertices
\vertex[above=0.05cm of ch1] (p1);
\vertex[below=0.05cm of ch1] (p2);
\vertex[right=0.54cm of p1] (q1);
\vertex[right=0.46cm of p2] (q2);
\diagram* {
{
(a) -- [fermion, small] (b),
(b) -- (c1) -- (c2)  -- [fermion, small] (o2),
(b) -- (d1) -- (d2)  -- [anti fermion, small] (o3),
(c1) -- (d2),
(d1) -- (q2), (c2) -- (q1)
}
};
\end{feynman}
\end{tikzpicture}}&
\scalebox{1.18}{\begin{tikzpicture}
\begin{feynman}
\vertex (a) {$p^\prime$};
\vertex[right=1cm of a] (b);
\vertex[right=1cm of b] (ch1);
\vertex[above=0.7cm of ch1] (c1);
\vertex[above=0.5cm of c1] (o1);
\vertex[right=1cm of c1] (c2);
\vertex[right=0.5cm of c2] (o2) {$q-p_1$};
\vertex[below=0.7cm of ch1] (d1);
\vertex[right=1cm of d1] (d2);
\vertex[right=0.5cm of d2] (o3) {$p_2$};
%more auxiliary vertices
\vertex[above=0.05cm of ch1] (p1);
\vertex[below=0.05cm of ch1] (p2);
\vertex[right=0.54cm of p1] (q1);
\vertex[right=0.46cm of p2] (q2);
\diagram* {
{
(a) -- [fermion, small] (b),
(b) -- (c1) -- (c2)  -- [anti fermion, small] (o2),
(b) -- (d1) -- (d2)  -- [fermion, small] (o3),
(c1) -- (d2),
(d1) -- (q2), (c2) -- (q1)
}
};
\end{feynman}
\end{tikzpicture}}\\[2.0em]
\scalebox{1.30}{\feynmandiagram [layered layout, small, horizontal=a to b] {
% Draw the top and bottom lines
i1 [particle=\(q\)]
--  [fermion] a %[dot]
--  b %[dot]
--  [fermion] f1 [particle=\(p_2\)],
i2 [particle=\(p_1\)]
-- [anti fermion] d %[dot]
--  e %[dot]
-- [anti fermion] f2 [particle=\(p^\prime\)],
% Draw the two internal fermion lines
{ [same layer] a -- d },
{ [same layer] b -- e},
{ [same layer] b -- d},
};}&
\scalebox{1.18}{\begin{tikzpicture}
\begin{feynman}
\vertex (a) {$p_1$};
\vertex[right=1cm of a] (b);
\vertex[right=1cm of b] (ch1);
\vertex[above=0.7cm of ch1] (c1);
\vertex[right=0.5cm of c1] (o2) {$q$};
\vertex[below=0.7cm of ch1] (d1);
\vertex[right=0.5cm of d1] (o3) {$p^\prime-p_2$};
\diagram* {
{
(a) -- [anti fermion, small] (b),
(b) -- (c1) -- [anti fermion, small] (o2),
(b) -- (d1) -- [anti fermion, small] (o3),
(b) -- (ch1),
(c1) -- (d1)
}
};
\end{feynman}
\end{tikzpicture}}&
\scalebox{1.18}{\begin{tikzpicture}
\begin{feynman}
\vertex (a) {$p^\prime - p_2$};
\vertex[right=1cm of a] (w);
\vertex[right=0.7cm of w] (c);
\vertex[right=0.7cm of c] (e);
\vertex[right=0.5cm of e] (o2) {$p_1$};
\vertex[above=0.7cm of c] (n);
\vertex[above=0.5cm of n] (o1);
\vertex[below=0.7cm of c] (s);
\diagram* {
{
(a) -- [fermion, small] (w),
(w) -- [quarter left] (n) -- [quarter left] (e) -- [quarter left] (s) -- [quarter left] (w) ,
(w) -- (c) -- (e) -- [fermion, small] (o2),
(n) -- [anti fermion, small, edge label=\(q\)] (o1)
}
};
\end{feynman}
\end{tikzpicture}}&
\scalebox{1.18}{\begin{tikzpicture}
\begin{feynman}
\vertex (a) {$p^\prime$};
\vertex[right=0.7cm of a] (w);
\vertex[right=0.7cm of w] (c);
\vertex[right=0.7cm of c] (e);
\vertex[right=0.5cm of e] (o2) {$p_1 + p_2$};
\vertex[above=0.7cm of c] (n);
\vertex[above=0.5cm of n] (o1);
\vertex[below=0.7cm of c] (s);
\diagram* {
{
(a) -- [fermion, small] (w),
(w) -- [quarter left] (n) -- [quarter left] (e) -- [quarter left] (s) -- [quarter left] (w) ,
(w) -- (c) -- (e) -- [fermion, small] (o2),
(n) -- [anti fermion, small, edge label=\(q\)] (o1)
}
};
\end{feynman}
\end{tikzpicture}}\\[2.0em]
\scalebox{1.18}{\begin{tikzpicture}
\begin{feynman}
\vertex (a) {$q$};
\vertex[right=0.7cm of a] (w);
\vertex[right=0.7cm of w] (c);
\vertex[right=0.7cm of c] (e);
\vertex[right=0.5cm of e] (o2) {$p_1$};
\vertex[above=0.7cm of c] (n);
\vertex[above=0.5cm of n] (o1);
\vertex[below=0.7cm of c] (s);
\diagram* {
{
(a) -- [fermion, small] (w),
(w) -- [quarter left] (n) -- [quarter left] (e) -- [quarter left] (s) -- [quarter left] (w) ,
(w) -- (c) -- (e) -- [fermion, small] (o2),
(n) -- [anti fermion, small, edge label=\(p^\prime - p_2\)] (o1)
}
};
\end{feynman}
\end{tikzpicture}}&
\scalebox{1.18}{\begin{tikzpicture}
\begin{feynman}
\vertex (a) {$p^\prime$};
\vertex[right=0.7cm of a] (w);
\vertex[right=0.7cm of w] (c);
\vertex[right=0.7cm of c] (e);
\vertex[right=0.5cm of e] (o2) {$p_1$};
\vertex[above=0.7cm of c] (n);
\vertex[above=0.5cm of n] (o1);
\vertex[below=0.7cm of c] (s);
\diagram* {
{
(a) -- [fermion, small] (w),
(w) -- [quarter left] (n) -- [quarter left] (e) -- [quarter left] (s) -- [quarter left] (w) ,
(w) -- (c) -- (e) -- [fermion, small] (o2),
(n) -- [anti fermion, small, edge label=\(p_1 - p^\prime\)] (o1)
}
};
\end{feynman}
\end{tikzpicture}}&
\scalebox{1.18}{\begin{tikzpicture}
\begin{feynman}
\vertex (a) {$P$};
\vertex[right=0.7cm of a] (w);
\vertex[right=0.7cm of w] (c);
\vertex[right=0.7cm of c] (e);
\vertex[right=0.5cm of e] (o2);
\vertex[above=0.7cm of c] (n);
\vertex[below=0.7cm of c] (s);
\diagram* {
{
(a) -- [fermion, small] (w),
(w) -- [quarter left] (n) -- [quarter left] (e) -- [quarter left] (s) -- [quarter left] (w) ,
(w) -- (c) -- (e) -- [fermion, small] (o2)
}
};
\end{feynman}
\end{tikzpicture}}&
\scalebox{1.18}{\begin{tikzpicture}
\begin{feynman}
\vertex (a) {$P$};
\vertex[right=0.7cm of a] (w);
\vertex[right=0.7cm of w] (c);
\vertex[right=0.7cm of c] (e);
\vertex[right=0.5cm of e] (o2);
\vertex[above=0.7cm of c] (n);
\vertex[below=0.7cm of c] (s);
\diagram* {
{
(a) -- [fermion, small] (w),
(w) -- [quarter left] (n) -- [quarter left] (e) -- [quarter left] (s) -- [quarter left] (w) ,
(e) -- [fermion, small] (o2)
}
};
\end{feynman}
\end{tikzpicture}}
\end{tabular}}
}
\caption{Complete set of master integrals for the two-loop calculation.\label{fig:masters}}
\end{figure}
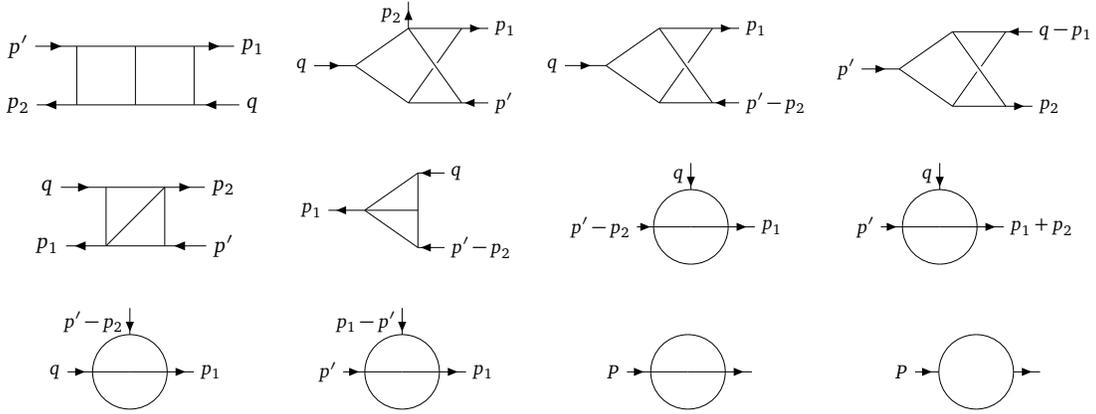

In total, we get the $12$ independent master integrals depicted in Fig.~\ref{fig:masters}. The easier ones among them
can be solved in closed form in terms of Gamma- and hypergeometric functions,
which we expand in $\epsilon$ with {\tt HypExp}~\cite{Huber:2005yg,Huber:2007dx}.
The more complicated ones are evaluated with the method of differential equations~\cite{Kotikov:1990kg,Kotikov:1991pm,Remiddi:1997ny,Argeri:2007up},
partially in a canonical basis~\cite{Henn:2013pwa}.
Furthermore, Mellin-Barnes representations~\cite{Czakon:2005rk} are employed,
both to compute asymptotic expansions for $x \to 0$ or $x \to 1$ to get the boundary conditions for the differential equations,
and to derive the full %$x$-dependence of 
analytic expressions of the master integrals.
The sector decomposition program {\tt FIESTA}~\cite{Smirnov:2008py,Smirnov:2015mct} is used to perform numerical checks of our analytic results.
We obtain the $\epsilon$-expansion of all master integrals analytically
in terms of harmonic polylogarithms (HPLs)~\cite{Remiddi:1999ew,Maitre:2005uu,Maitre:2007kp} with weight-indices $0$~and $1$.
To the order we are working, HPLs of at most weight four appear in the amplitude.
Below, we give the results for the master integrals in the first column of Fig.~\ref{fig:masters}, which we label $I_1(x)$, $I_5(x)$ and $I_9(x)$. Using $\int d^Dk/(2\pi)^D$ as integration measure, and factoring out $(i/((4\pi)^{D/2}\Gamma(1-\epsilon)))^2$ together with an appropriate power of $Q^2$ to make the integral dimensionless, we obtain ($H_{a_1,\ldots,a_n}(x)$ are HPLs)
\allowdisplaybreaks
\begin{align}
\bar x^2 x \epsilon^4 \, I_1(x) &= -1+2\epsilon (H_{0}(x)-H_{1}(x))+\epsilon^2 \left(H_{2}(x)-4 H_{0,0}(x)+3 H_{1,0}(x)-4 H_{1,1}(x)-\textstyle \frac{1}{2}\displaystyle\zeta_2\right) \nonumber \\[0.4em]
&+\epsilon^3 \left(\zeta_2 H_{0}(x)-3 \zeta_2 H_{1}(x)-2 H_{3}(x)+2 H_{1,2}(x)+H_{2,0}(x)-H_{2,1}(x)+8 H_{0,0,0}(x) \right.\nonumber \\[0.4em]
&\left.-6 H_{1,0,0}(x)+4 H_{1,1,0}(x)-8 H_{1,1,1}(x)+\textstyle\frac{13}{2}\displaystyle \zeta_3\right)
+\epsilon^4 \left(-13 \zeta_3 H_{0}(x)+5 \zeta_3 H_{1}(x) \right.\nonumber \\[0.4em]
&-\zeta_2  H_{2}(x)-2\zeta_2  H_{0,0}(x)-10\zeta_2 H_{1,1}(x)-2 H_{1,3}(x)-5 H_{2,1,1}(x)-2 H_{2,0,0}(x)\nonumber \\[0.4em]
&-2 H_{3,0}(x)-H_{3,1}(x)+4 H_{1,1,2}(x)+4 H_{4}(x)-H_{2,1,0}(x)-H_{2,2}(x)-8 H_{1,1,0,0}(x)\nonumber \\[0.4em]
&\left.-16 H_{0,0,0,0}(x)+12 H_{1,0,0,0}(x)+4 H_{1,1,1,0}(x)-16 H_{1,1,1,1}(x)\right) + {\cal O}(\epsilon^5) \, ,\\[0.4em]
\epsilon^4 \, I_5(x) &= \epsilon^3\left(-\zeta_2 H_{0}(x)-\zeta_2 H_{1}(x)+H_{3}(x)+H_{1,2}(x)-H_{2,0}(x)-H_{1,1,0}(x)+3 \zeta_3\right)\nonumber \\[0.4em]
&+\epsilon^4 \left(-7 \zeta_3 H_{0}(x)+5 \zeta_3 H_{1}(x)+\zeta_2 H_{2}(x)-2 H_{1,2,0}(x)+2\zeta_2 H_{0,0}(x)-2\zeta_2 H_{1,0}(x)\right.\nonumber \\[0.4em]
&-3\zeta_2 H_{1,1}(x)+\! H_{2,1,0}(x)-2 H_{2,2}(x)+\! 2 H_{3,0}(x)+2 H_{3,1}(x)+2 H_{1,1,2}(x)-3 H_{4}(x)\nonumber \\[0.4em]
&\left.+2 H_{1,2,1}(x)+2 H_{2,0,0}(x)+H_{1,3}(x)+2 H_{1,1,0,0}(x)-3 H_{1,1,1,0}(x)+\textstyle\frac{51}{4}\zeta_4\displaystyle\right)+ {\cal O}(\epsilon^5) \, ,\\[0.4em]
I_9(x) &= \frac{\Gamma^5(1-\epsilon)\Gamma (1-2 \epsilon)  \Gamma (\epsilon) \Gamma (2 \epsilon) }{\Gamma (2-3 \epsilon) \Gamma (2-2 \epsilon) \Gamma (1+\epsilon)} \, {}_2F_1(1,2\epsilon;1+\epsilon;x) \, .
\end{align}

%%%%%%%%%%%%%%%%%%%%%%%%%%%%%%%%%%%%%%%%%%%%%%%%%%%%%%%%%%%%%%%%%%%%%%%%%%%%%%%%%%%%%%%%%%%%%%%%%%%%
%%%%%%%%%%%%%%%%%%%%%%%%%%%%%%%%%%%%%%%%%%%%%%%%%%%%%%%%%%%%%%%%%%%%%%%%%%%%%%%%%%%%%%%%%%%%%%%%%%%%

\section{UV renormalization and IR subtraction}
\label{sec:uvir}

To arrive at the two-loop contribution to $T_2(x)$ we must perform UV renormalization and IR subtraction.
This is done by first introducing the basis $\{ O_{1}^{\mu \nu}, \, O_{2}^{\mu \nu}, \, O_{E}^{\mu \nu} \}$ of non-local operators,
\begin{align}
O_{j}^{\mu \nu}(x)= \frac{\bar n \cdot p}{2 \pi} \int d \tau  \,
 \e^{{ i \bar x  \tau \bar n \cdot p}}
\bar q (\tau \bar n) \, [\tau \bar n,0]\,  \Gamma_{j}^{\mu \nu}  \,  q(0) \,,
\end{align}
 where
\begin{align}
\Gamma_{1}^{\mu \nu} &= g_{\perp}^{\mu \nu}  \, \slashed{\bar n}  \,, \qquad
\Gamma_{2}^{\mu \nu} =  i \, \epsilon_{\perp}^{\mu \nu}  \, \slashed{\bar n} \,  \gamma_5 \,, \qquad
\Gamma_{E}^{\mu \nu} = \slashed{\bar n}
\left ( {\textstyle\frac{1}{2}\displaystyle [ \gamma_{\perp}^{\mu},  \gamma_{\perp}^{\nu}] }
-  i \, \epsilon_{\perp}^{\mu \nu} \,  \gamma_5 \right ) \,,
\end{align}
and then by exploiting the matching equation
\begin{align}
\Pi^{\mu\nu} = \sum\limits_{a=1,2,E} T_a \otimes \langle O_{a}^{\mu\nu} \rangle\, .
\label{eq:matching}
\end{align}
By employing the definitions  $g_{\perp}^{\mu \nu} = g^{\mu \nu}-n^{\mu} \bar n^{\nu}/2 -n^{\nu} \bar n^{\mu}$/2
and $\epsilon^{\mu \nu}_{\perp}  \equiv \epsilon^{\mu \nu \alpha \beta} \bar n_{\alpha} \, n_{\beta}/2$, 
we see that $O_{E}^{\mu \nu}$ is an evanescent operator, i.e.\ it vanishes at $D=4$. Since furthermore
the CP-even operator $O_{1}^{\mu \nu}$ cannot couple to the pseudoscalar $\pi^{0}$ state, we encounter a unique physical operator $O_{2}^{\mu \nu}$~\cite{Wang:2017ijn}. Moreover, the Dirac structures are related via the algebraic identities
$\Gamma_{A,B}^{\mu \nu} = - (\Gamma_{1}^{\mu \nu} \pm \Gamma_{2}^{\mu \nu}\pm  \Gamma_{E}^{\mu \nu})$
which we use to switch between different notations.

The correlator $\Pi^{\mu\nu}$ assumes the following form to all orders in $\alpha_s$ in terms of the tree-level matrix elements,
\begin{align}
    \Pi^{\mu\nu}= \!\!\sum_{k=1,2,E}\, \sum_{\ell=0}^\infty  (Z_\alpha  a_s)^\ell\,  A_k^{(\ell)}(x)\,\vev{\bar q(\bar xp)\Gamma_{k}^{\mu\nu}q(xp)}^{(0)} 
    \, , \label{eq:correlator}
\end{align}
where the coupling renormalization factor is given by  $Z_\alpha = 1 - a_s \beta_0/\epsilon + {\cal O}(a_s^2)$.
Hereafter we disregard $O_1^{\mu\nu}$ completely from our consierations due to parity. The functions $T_a$, $a=\{2,E\}$ are expanded as in Eq.~(\ref{eq:loopexpansion}). Due to scaleless integrals in dimensional regularization the matrix elements of the light-cone operators are simply expanded as
\begin{align}
\vev{O_a^{\mu\nu}} 
= & \sum_{\ell=0}^\infty a_s^\ell \, Z_{ab}^{(\ell)} \otimes \vev{O_b^{\mu\nu}}^{(0)} 
 =  \left\{ \delta_{ab} + a_s \, Z_{ab}^{(1)} +  a_s^2 \, Z_{ab}^{(2)} + {\cal O}(a_s^3)\right\} \otimes \langle O_b^{\mu\nu} \rangle^{(0)} \, .
\label{eq:MEOa}
\end{align}
As indicated above, sums over repeated indices run over $\{2,E\}$. $\langle O_{a}^{\mu\nu} \rangle^{(0)}$ is the tree-level matrix element of $O_a^{\mu\nu}$, and the renormalization constants $Z_{22}^{(\ell)}$ can be extracted from the Efremov-Radyushkin-Brodsky-Lepage (ERBL) kernel~\cite{Efremov:1979qk, Lepage:1980fj,Sarmadi:1982yg,Dittes:1983dy,Katz:1984gf,Mikhailov:1984ii,Belitsky:1999gu}.
Inserting Eqs.~\eqref{eq:loopexpansion},~\eqref{eq:correlator}, and~\eqref{eq:MEOa} into Eq.~\eqref{eq:matching} and comparing coefficients of $\langle O_{2,E}^{\mu\nu} \rangle^{(0)}$ at any given order in $a_s$, we derive the following ``master formulas'' for the hard CF at the various loop orders
\begin{align}
T_2^{(0)}&=T_E^{(0)}=A_2^{(0)}\, ,\notag\\[0.3em]
T_2^{(1)}&= A_2^{(1)} - \sum_{a=2,E} Z_{a2}^{(1)} \otimes T_a^{(0)} =T_E^{(1)}-Z_{E2}^{(1)} \otimes T_E^{(0)} \, ,\notag\\
T_2^{(2)} &= A_2^{(2)} + Z_\alpha^{(1)} A_2^{(1)} - \sum_{a=2,E}\sum_{k=0}^1 Z_{a2}^{(2-k)}\otimes T_a^{(k)}  \, .
\label{eq:masterformula}
\end{align}
The derivation of these formulas made further use of the relations $A_2^{(\ell)} = A_E^{(\ell)}$, $Z_{EE}^{(1)} = Z_{22}^{(1)}$ and $Z_{12}^{(1)}=Z_{2E}^{(1)}=0$. 

Finally, a word on the mixing between evanescent and physical operators is in order. The master formulas~\eqref{eq:masterformula} were derived under the assumption that dimensional regularization is in Eq.~\eqref{eq:matching} for both UV and IR divergences. However, to determine the UV-renormalization constants $Z_{ab}^{(\ell)}$ a different procedure has to be adopted. To this end, the renormalized matrix elements of the effective operators are expressed as 
\begin{align}
 \langle O_{a}^{\mu\nu} \rangle  = & \, \left\{ \delta_{ab} + a_s \, \left[M_{ab}^{(1)} + Z_{ab}^{(1)}\right] +  a_s^2 \, \left[M_{ab}^{(2)} + Z_{ab}^{(2)} + Z_{a2}^{(1)} \otimes M_{2b}^{(1)}  \right] + {\cal O}(a_s^3)\right\} \otimes \langle O_b^{\mu\nu} \rangle^{(0)} \, ,
\end{align}
and the matrix elements $M_{ab}^{(\ell)}$ are obtained with dimensional regularization applied only to the UV divergences  but with the IR regularization scheme being different from the dimensional one, see e.g.\ also~\cite{Beneke:2009ek,Wang:2017ijn}. Renormalizing the matrix elements of the evanescent operator to zero yields the relations
\begin{align}
Z_{E2}^{(1)} & = - M_{E2}^{(1)} \, , &
Z_{E2}^{(2)} & = - M_{E2}^{(2)} + M_{E2}^{(1)} \otimes M_{22}^{(1)} \, ,
\end{align}
where $Z_{E2}^{(2)}$ is IR finite albeit both $M_{E2}^{(2)}$ and $M_{22}^{(1)}$ being IR divergent.

Collecting all individual pieces in \eqref{eq:masterformula} together, all the poles in $\epsilon$ are canceled explicitly as expected, which represents a nontrivial check for our calculation. As stated earlier, our final analytic result for the hard CF $T_2^{(2)}(x)$ agrees with that of a parallel computation~\cite{Braun:2021grd}, which uses quite a different approach based on arguments from conformal symmetry. The result for $T_2^{(2)}(x)$ is split up into three colour factors and can be expressed as
\begin{align}
T_2^{(2)}(x) &=  \, \beta_0 \, C_F \left( {\cal K}_{\beta}^{(2)}(x) / x + {\cal K}_{\beta}^{(2)}(\bar x) / \bar x \right)
             + C_F^2 \left({\cal K}_{F}^{(2)}(x) / x + {\cal K}_{F}^{(2)}(\bar x) / \bar x \right) \nonumber \\
	     &+ C_F / N_c \left({\cal K}_{N}^{(2)}(x) / x + {\cal K}_{N}^{(2)}(\bar x) / \bar x \right) \, .\label{eq:T22-expr}
\end{align}
Our result for the color structure $C_F \beta_0$ agrees with the NNLO computation of the large-$\beta_0$ limit in~\cite{Melic:2001wb}. The explicit expressions of the functions ${\cal K}_{\beta,F,N}^{(2)}(x)$ are too lengthy to be given explicitly here. They can be found in~\cite{Gao:2021iqq}, whose arXiv repository also contains the functions in electronic form.

\section{Numerical results}
\label{sec:results}

To facilitate our numerical analysis, certain assumptions (models) on the non-perturbative twist-two pion LCDA $\phi_\pi(x,\mu_0)$ have to be made. 
It is convenient to construct/constrain the phenomenological models via the expansion in Gegenbauer polynomials,  i.e.,
\begin{align}
   \phi_\pi(x,\mu_0)&=6x\bar x\left(1+\sum_{n=1}^\infty a_{2n}(\mu_0)\right)\, C_{2n}^{3/2}(x-\bar x)\, ,
\end{align}
where $a_{2n}(\mu_0)$ are called Gegenbauer moments. 
It is customary to set the reference scale $\mu_0=1~{\rm GeV}$ for model building and then evolve $\phi_\pi(x,\mu_0)$ to the proper factorization scale $\mu_F$ with the help of renormalization group equation (RGE) where the one-loop, two-loop, and three-loop ERBL evolution kernels (anomalous dimension matrix) are available in~\cite{Efremov:1979qk, Lepage:1980fj},~\cite{Sarmadi:1982yg,Dittes:1983dy,Katz:1984gf,Mikhailov:1984ii,Belitsky:1999gu}, and~\cite{Braun:2017cih}, respectively. 

\begin{figure}[tp]
\begin{center}
\includegraphics[width=0.6 \columnwidth]{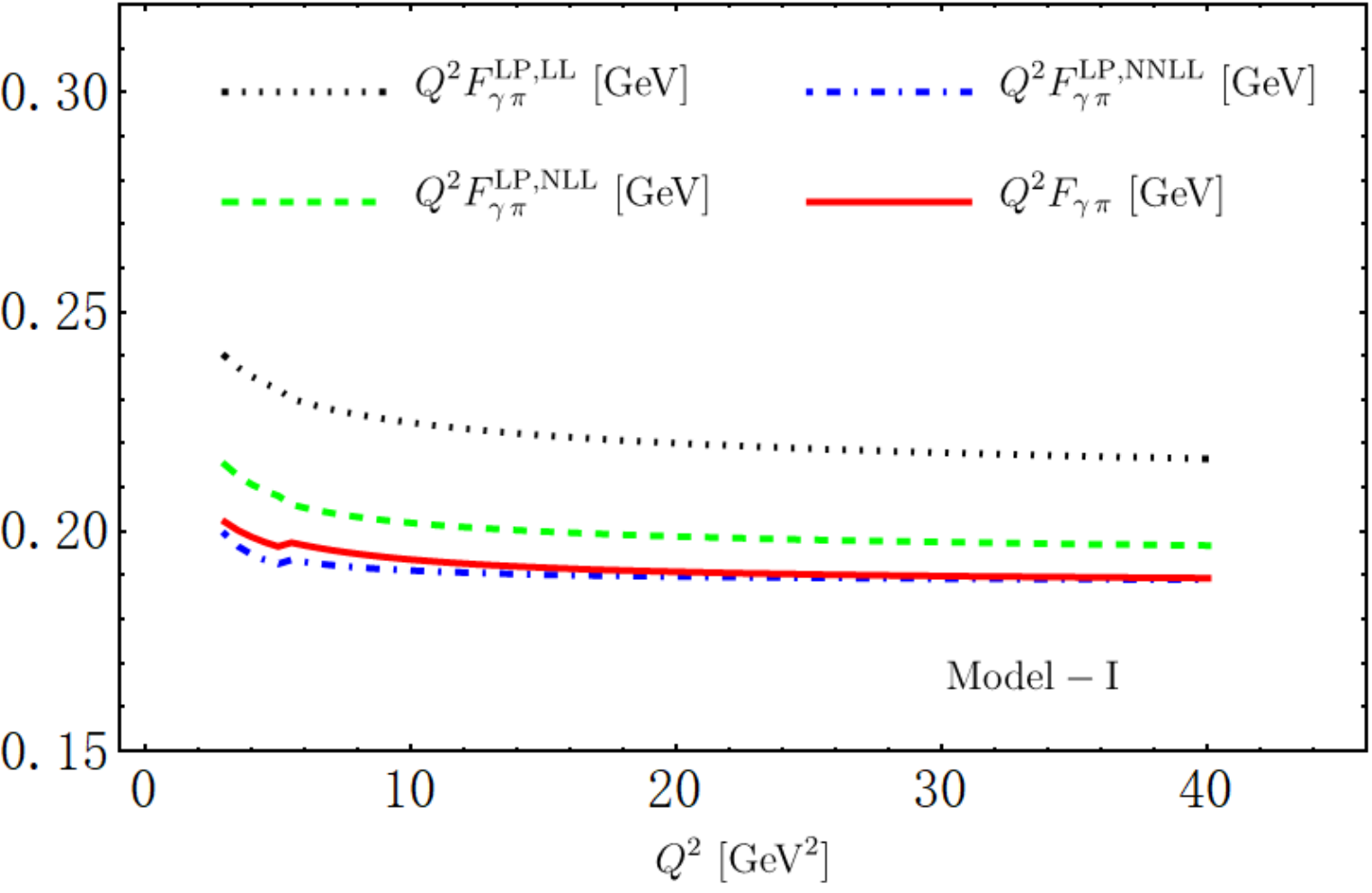}
\vspace*{0.1cm}
\caption{Theory predictions of $F_{\gamma\pi}(Q^2)$  
at the LL (black dotted), NLL (green dashed), and NNLL (blue dot-dashed) order %[black dotted],
%at the next-to-leading-logarithmic (NLL) order [green dashed],
%and at the NNLL order [blue dot-dashed] 
in QCD, respectively, adopting Model I in Eq.~\eqref{Models of the pion DA} as the non-perturbative input for $\phi_\pi(x,\mu_0)$. %accommodating the recent lattice result~\cite{RQCD:2019osh}.
The red solid curve further includes 
the subleading power contributions evaluated in \cite{Khodjamirian:1997tk,Wang:2017ijn}.}
\label{fig: Breakdown of the pion FF for Model-I}
\end{center}
\end{figure}

As an illustration of the two-loop effect on the observable $F_{\gamma\pi}(Q^2)$, we adopt the so-called ADS/QCD model for $\phi_\pi(x,\mu_0)$ proposed in~\cite{Brodsky:2007hb} with modifications from recent lattice computation of the first nontrivial Gegenbauer moment $a_2(\mu_0)$~\cite{RQCD:2019osh} (Model I in Eq.~\eqref{Models of the pion DA} below). 
We display the numerical impact of QCD corrections on $F_{\gamma\pi}(Q^2)$  in Fig.~\ref{fig: Breakdown of the pion FF for Model-I} %following the standard nomenclature, 
at leading-logarithmic order (LL), next-to-leading-logarithmic order (NLL), and next-to-next-to-leading-logarithmic order (NNLL), 
by including the renormalization-group evolution effect of the leading-twist pion LCDA  at one loop, two loops, and three loops.
%indicating that the $a_s^n$ and $a_s^{n+1}$ corrections to $T_2$ and $\phi_\pi(x,\mu)$ are included at the ${\rm N}^n{\rm LL}$ order, respectively. 
%
%The necessity of taking the $a_s$ correction to $\phi_\pi(x,\mu)$ one order higher than $T_2(x)$ %due to numerically $\log (Q^2/ \Lambda_{\rm QCD}^2) \sim 1/\alpha_s(Q)$
%allows for proper resummation of large logarithms and hence provides us with more control over numerical predictions.
%

%%%%%%%%%%%%%%%%%%%%%%%%%%%%%%%%%%%%%%%%%%%%%%%%%%%%%%%%%%%%%%%
%%%%%%%%%%%%%%%%%%%%%%%%%%%%%%%%%%%%%%%%%%%%%%%%%%%%%%%%%%%%%%%
%

It is clear from Fig.~\ref{fig: Breakdown of the pion FF for Model-I} that the two-loop correction to the photon-pion TFF is rather significant. 
To be more specific, the NNLL correction is responsible for a $(4\sim7)\%$ reduction to $F^{\rm LP,NLL}_{\gamma\pi}(Q^2)$ in the kinematic region $Q^2\in[3,40]~{\rm GeV}^2$, in comparison to the $\sim 10\%$ reduction when advancing from $F^{\rm LP,LL}_{\gamma\pi}$ to $F^{\rm LP,NLL}_{\gamma\pi}$ accuracy.
This pattern of QCD correction hierarchy is also observed for the other two representative models presented in Eq.~\eqref{Models of the pion DA} which validates the significance of our two-loop computation in general.

Let us now discuss the potentiality of our theory predictions in unraveling the omnipresent non-perturbative object $\phi_\pi(x,\mu)$. 
For this purpose, we consider three representative models,
\begin{eqnarray}
{\rm Model \,\, I:} \,\,\, && \phi_{\pi}(x, \mu_0) =
\frac{\Gamma(2 + 2 \alpha_{\pi})}{  \Gamma^2(1 + \alpha_{\pi})}  \,
(x \, \bar x)^{\alpha_{\pi}} \,,\quad
 {\rm with } \,\,\quad \alpha_{\pi}(\mu_0) = 0.422^{+0.076}_{-0.067}  \,;
\label{Models of the pion DA}  \\[0.6em]
{\rm Model \,\, II:} \,\,\, && \{a_{2},a_{4}, a_6, a_8\}(\mu_0) = \{0.269(47),\, 0.185(62),\, 0.141(96),\, 0.049(116)\} \,;
\nonumber \\[0.6em]
{\rm Model \,\, III:} \,\,\, && \{ a_{2},a_{4} \}(\mu_0) = \{0.203^{+0.069}_{-0.057} \,,
-0.143^{+0.094}_{-0.087}\} \,.
\nonumber
\end{eqnarray}
While the background of Model I has been elaborated on above, Model II~\cite{Cheng:2020vwr} and III~\cite{Bakulev:2001pa,Mikhailov:2016klg,Stefanis:2020rnd} are proposed from the perspective of light-cone and QCD sum rules, respectively.  
%
%
%
%%%%%%%%%%%%%%%%%%%%%%%%%%%%%%%%%%%%%%%%%%%%%%%%%%%%%%%%%%%%%%%
\begin{figure}[tp]
\begin{center}
\includegraphics[width=0.6 \columnwidth]{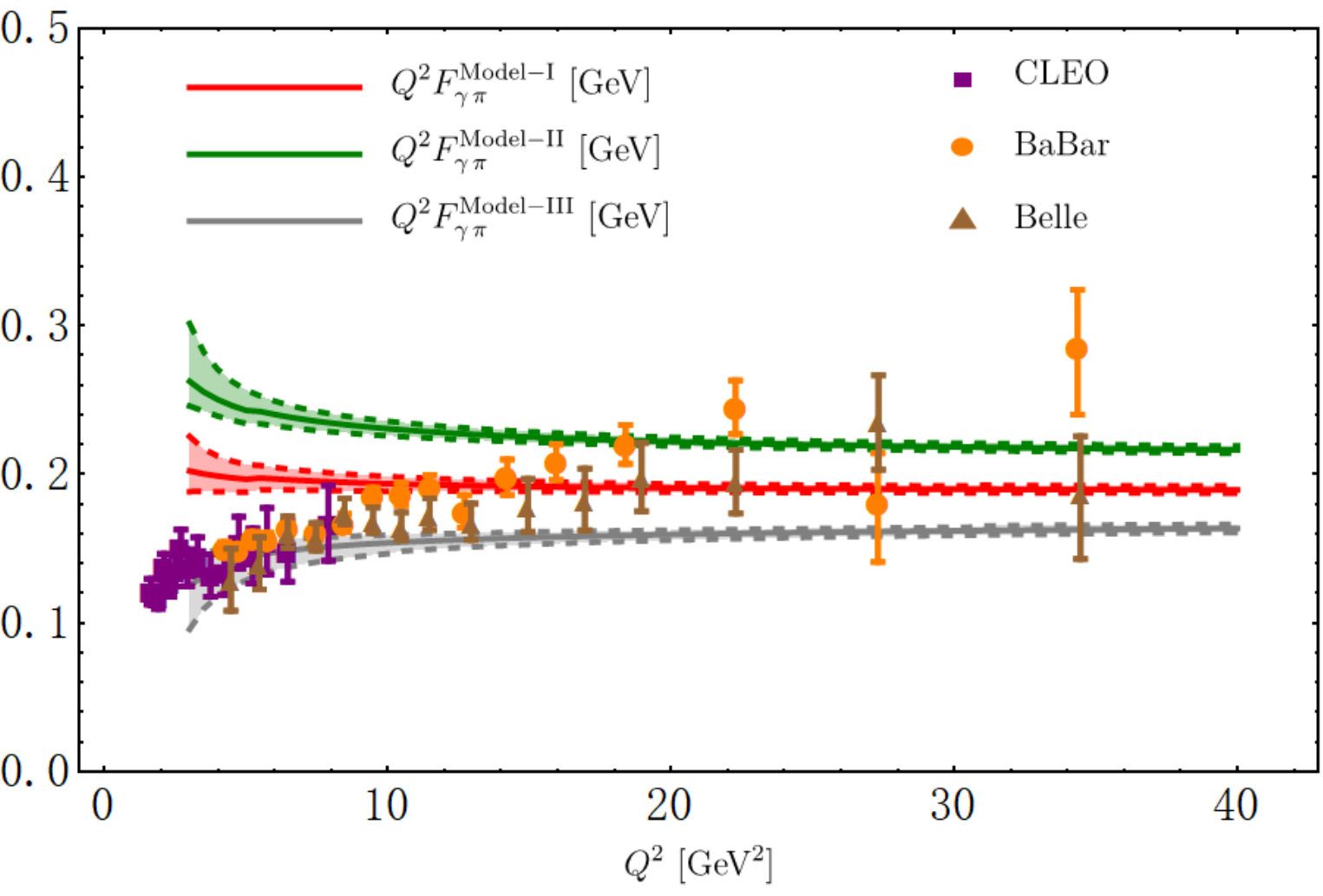}
\vspace*{0.1cm}
\caption{Theory predictions of the TFF
$\gamma\gamma^{\ast} \to \pi^0$ based on the three models 
in~\eqref{Models of the pion DA}.
The color bands indicate uncertainties from factorization/renormalization scale variation $\mu_F\in [1/4,3/4]Q^2$.
For a comparison, our predictions are confronted with the experimental measurements
from the CLEO~\cite{Gronberg:1997fj} (purple squares),
BaBar~\cite{Aubert:2009mc} (orange circles)
and Belle \cite{Uehara:2012ag} (brown spades) Collaborations.}
\label{fig: Model dependence of the pion FF}
\end{center}
\end{figure}
%%%%%%%%%%%%%%%%%%%%%%%%%%%%%%%%%%%%%%%%%%%%%%%%%%%%%%%%%%%%%%%
%
%
%

We collect our theory predictions of $F_{\gamma\pi}(Q^2)$ for the three models in~\eqref{Models of the pion DA} in Fig.~\ref{fig: Model dependence of the pion FF}, where we have purposefully neglected the error-bars quoted in Eq.~\eqref{Models of the pion DA} to enhance the characteristics of each model.
It is then evident that with our two-loop correction taken into account, the perturbative uncertainties are small enough to allow to distinguish various phenomenological models for $\phi_\pi(x,\mu_0)$.
As these models are formulated on vastly different principles, a concrete determination of an appropriate description for $\phi_\pi(x,\mu)$ from future experimental data will certainly be crucial to deepen our understanding of the inner structures of the hadronic states.

%%%%%%%%%%%%%%%%%%%%%%%%%%%%%%%%%%%%%%%%%%%%%%%%%%%%%%%%%%%%%%%%%%%%%%%%%%%%%%%%%%%%%%%%%%%%%%%%%%%%
%%%%%%%%%%%%%%%%%%%%%%%%%%%%%%%%%%%%%%%%%%%%%%%%%%%%%%%%%%%%%%%%%%%%%%%%%%%%%%%%%%%%%%%%%%%%%%%%%%%%

\section{Conclusion}

In conclusion, we have promoted the leading-power theory prediction of the photon-pion transition form factor to the two-loop order in QCD by combining hard-collinear factorization with modern multi-loop techniques. The analytic two-loop coefficient function we have obtained is also directly applicable in the evaluation of the axial-vector contribution to  deeply virtual Compton scattering. In this article, we have presented the complete set of two-loop master integrals arising in the calculation of $\gamma\gamma^*$ annihilation into two collinear  on-shell massless quarks.  We expect our results to play an integral role in boosting future developments on the determination of the leading-twist pion LCDA as well as on exploring  the delicate QCD dynamics of other interesting two-photon processes.

%%%%%%%%%%%%%%%%%%%%%%%%%%%%%%%%%%%%%%%%%%%%%%%%%%%%%%%%%%%%%%%%%%%%%%%%%%%%%%%%%%%%%%%%%%%%%%%%%%%%
%%%%%%%%%%%%%%%%%%%%%%%%%%%%%%%%%%%%%%%%%%%%%%%%%%%%%%%%%%%%%%%%%%%%%%%%%%%%%%%%%%%%%%%%%%%%%%%%%%%%

\section*{Acknowledgements}
We would like to thank the authors of~\cite{Braun:2021grd} for useful correspondence, and for sharing their analytic results with us prior to publication. 

% TODO: include funding information
\paragraph{Funding information}
J.G.~was partially supported by the Deutscher Akademischer Austauschdienst (DAAD).
The research of T.H.~and Y.J.~was supported in part by the Deutsche Forschungsgemeinschaft
(DFG, German Research Foundation) under grant  396021762 - TRR 257 (``Particle Physics Phenomenology after the Higgs Discovery'') and DFG grant SFB TR 110/2.
Y.M.W.~acknowledges support from the National Youth Thousand Talents Program,
the Youth Hundred Academic Leaders Program of Nankai University,
the  National Natural Science Foundation of China  with Grant No. 11675082, 11735010 and 12075125,
and  the Natural Science Foundation of Tianjin with Grant No. 19JCJQJC61100.

% TODO:
% Provide your bibliography here. You have two options:

% FIRST OPTION - write your entries here directly, following the example below, including Author(s), Title, Journal Ref. with year in parentheses at the end, followed by the DOI number.
%\begin{thebibliography}{99}
%\bibitem{1931_Bethe_ZP_71} H. A. Bethe, {\it Zur Theorie der Metalle. i. Eigenwerte und Eigenfunktionen der linearen Atomkette}, Zeit. f{\"u}r Phys. {\bf 71}, 205 (1931), \doi{10.1007\%2FBF01341708}.
%\bibitem{arXiv:1108.2700} P. Ginsparg, {\it It was twenty years ago today... }, \url{http://arxiv.org/abs/1108.2700}.
%\end{thebibliography}

% \bibliographystyle{SciPost_bibstyle} % Include this style file here only if you are not using our template
\bibliography{References.bib}

\nolinenumbers

\end{document}